# Algebraic skin effect in two-dimensional non-Hermitian metamaterials

Mingyang Li, Jing Lin*, and Kun Ding†

*Department of Physics, State Key Laboratory of Surface Physics, and Key Laboratory of Micro and Nano Photonic Structures (Ministry of Education), Fudan University, Shanghai 200438, China*

*Email: jing_lin@fudan.edu.cn, †E-mail: kunding@fudan.edu.cn

**Abstract**

Metamaterials have unlocked unprecedented control over light by leveraging novel mechanisms to expand their functionality. Non-Hermitian physics further enhances the tunability of non-Hermitian metamaterials (NHMs) through phenomena such as the non-Hermitian skin effect (NHSE), enabling applications like directional amplification. The higher-dimensional NHSE manifests unique effects, including the algebraic skin effect (ASE), which features power-law decay instead of exponential localization, allowing for quasi-long-range interactions. In this work, we establish apparent criteria for achieving ASE in two-dimensional reciprocal NHMs with anisotropic and complex dielectric tensors. By numerically and theoretically demonstrating ASE through mismatched optical axes and geometric structures, we reveal that ASE is governed by a generalized Fermi surface whose dimensionality exceeds that of the Fermi surface. We further propose and validate a realistic photonic crystal design for ASE, which is experimentally accessible. Our recipe for ASE provides a versatile pathway for broader generalizations, including three-dimensional structures, synthetic dimensions, and other classical wave systems, paving the way for advancements in non-Hermitian photonics.

**Introduction**

Metamaterials have revolutionized wave manipulation, enabling unprecedented control over light through meticulously designed microstructures [1-8]. Beyond sophisticated fabrication and precise measurement techniques [9-11], the capability relies heavily on exploring novel mechanisms [12-30], such as steering wavefronts using gradient phases [12,13], lasing based on topological modes [14-19], and localization coming from Moire patterns [20,21]. Undoubtedly, recent progresses in non-Hermitian physics can add another knob to these platforms by utilizing nonreciprocity, loss, and gain [31-43]. The celebrated non-Hermitian skin effect (NHSE) of one-dimensional (1D) systems [44-50], where the bulk eigenstates localize at the boundaries under open boundary conditions (OBCs), roots in the extension of the Brillouin zone (BZ) to a generalized Brillouin zone (GBZ) [44-47]. This extension allows wavevectors to take complex values for bulk eigenstates under OBCs in the thermodynamic limit. Such theoretical insights facilitate the integration of NHSE into metamaterials, enabling applications like directional amplification [48-50]. From the perspective of metamaterial theory and design, it requires the generalization of effective medium theory (EMT) [51-54] and connection with GBZ to apply in non-Hermitian metamaterials (NHMs) [55-60], which will allow the harness of non-Hermitian degrees of freedom with precision and efficiency.

Beyond 1D systems, NHSE in higher dimensions has revealed even more intriguing and unique phenomena [61-71], such as the geometry-dependent skin effect (GDSE) [62-65], algebraic skin effect (ASE) [66,67], and so forth [68-71]. Unlike the exponential localization of NHSE in 1D systems, higher-dimensional skin modes may exhibit hybrid localization behaviors [66,68]. For instance, ASE features a power-law decay far from boundaries, leading to quasi-long-range interactions [67,69]. These higher-dimensional effects are closely tied to system geometry and boundary conditions (BCs), offering exciting possibilities for tunable and long-range wave manipulation. As such, realizing ASE in NHMs is worth pursuing. Specifically, for the sake of accessibility in the experiment and universality for further investigations, it is pivotal to clarify the criteria for ASE within the framework of continuum medium (CM).

In this work, we begin with the general properties of NHSE and ASE by examining the localization behaviors of eigenstates, reflecting their distinctions in the imaginary parts of wavevectors. Building on these insights, we propose an anisotropic NHM and demonstrate the occurrence of ASE by deploying various two-dimensional (2D) geometries. By discretizing the NHM into a lattice model (LM) and solving for the eigenstates, we confirm the presence of ASE and validate the wavevector-based considerations, leading to the concept of generalized Fermi surface (GFS) [66]. Hence, the criteria for ASE in NHMs have been clearly outlined, and we finally propose one realistic design based on photonic crystals (PhCs), in which ASE has been successfully showcased. Our work not only provides an experimentally accessible design for ASE but also irons out the difficulties that are unclear in the realization of ASE in NHMs.

**Results**

    **General Criteria and Properties of ASE.** Before discussing ASE, we revisit the 1D scenario shown in Figure 1a. Specifically, the non-Bloch Hamiltonian is generally expressed as a Laurent polynomial $H(\beta = e^{ik+\mu}) = \sum_{n=-m}^{m} t_n \beta^n$, where $m$ represents the maximum hopping range, $t_n$ denotes the hopping strengths, and $k$ ($\mu$) is real (imaginary) parts of wavevectors. The OBC spectrum $\sigma^o$ (the blue line in Figure 1b) is significantly different from the periodic boundary condition (PBC) spectrum $\sigma^p$ (the grey line in Figure 1b), indicating the occurrence of NHSE (Figure 1c). From a theoretical perspective, such a discrepancy requires acquiring $\sigma^o$ through unit-cell level calculations, which is precisely the core of the 1D GBZ theory [44-47]. In addition to the characteristic equation (ChE) $\det[H(\beta) - \omega \mathbb{I}] = 0$ ($\omega \in \mathbb{C}$), the non-Bloch wavevectors $\beta$ must satisfy a further constraint arising from the zero boundary conditions (ZBCs)

$$|\beta_m(\omega)| = |\beta_{m+1}(\omega)|. \qquad (1)$$

Physically speaking, eq 1 is the standing wave condition (SWC), which is equivalent to ZBCs in 1D non-Hermitian systems [45]. The ChE and SWC together give rise to $\sigma^o$, which generally is a line in the complex frequency plane (Figure 1b) because both $\omega$ and $\beta$ are complex numbers. According to eq 1, an arbitrary frequency $\omega_0 \in \sigma^o$ now possesses two values of $k$ with the same $\mu$, possibly forming standing waves to satisfy ZBCs, with a typical eigenstate $\psi_0$ shown in Figure 1c. The localization clearly exhibits the exponential decay behavior, characterized by $|\psi_0| \propto e^{\mu x}$. In other words, a single attenuation factor $\mu$ is sufficient to describe the localization of the eigenstates.

    The generalization to 2D and even higher-dimensional non-Hermitian systems appears straightforward because both ChE and ZBCs are universal. Figure 1d shows a representative geometric configuration of a 2D LM with the typical $\sigma^p$ (grey area) and $\sigma^o$ (blue area) displayed in Figure 1e. Both $\sigma^p$ and $\sigma^o$ occupy areas instead of lines, and $\sigma^o \subset \sigma^p$ also hints at the presence of the NHSE. If we simply extend the conclusion from 1D systems, the

localization in higher dimensions can still be described by a single attenuation factor $\boldsymbol{\mu} = (\mu_x, \mu_y)$, indicating that the eigenstate decays exponentially along the $x$- and $y$-directions. This is illustrated by the frequency $\omega_1$ (black cross in Figure 1e) and its corresponding eigenstate $\psi_1$ (black lines in Figure 1f), which clearly localizes to a corner or edge determined solely by $\boldsymbol{\mu}$, independent of geometry. The higher-dimensional GBZ theory (or amoeba theory) has successfully determined $\boldsymbol{\mu}$ [72]. However, in reciprocal non-Hermitian systems, it is known that the NHSE can arise purely from geometric effects. In such systems, $\boldsymbol{\mu}$ determined by amoeba theory is always zero [72,73], suggesting that the eigenstate does not localize following the conventional sense of $\boldsymbol{\mu}$. This scenario is demonstrated by $\omega_2$ (red cross in Figure 1e) and its corresponding eigenstate $\psi_2$ (red lines in Figure 1f). Recent studies have shown that $\psi_2$ not only depends on the system geometry but also exhibits slower decay compared to $\psi_1$ (Figure 1f). Unlike $|\psi_1| \propto e^{\boldsymbol{\mu} \cdot \boldsymbol{r}}$, $|\psi_2|$ does not always follow an exponential decay but instead decays as a power law away from the boundary [66]. Based on the law of contraposition, this observation suggests that the localization for a class of higher-dimensional NHSEs is not from a single $\boldsymbol{\mu}$ but rather from a set of $\boldsymbol{\mu}$. This phenomenon leads to the celebrated ASE, which is distinct from the conventional NHSE. The quasi-long-range nature of ASE provides another knob to manipulate matter and waves, making it particularly worthy for realization in optics.

**CM to Realize ASE.** We commence from the CM to detail the realizing essentials. First and foremost, the eigenstate for a given frequency must localize according to the geometry rather than a single $\boldsymbol{\mu}$. It is the GDSE that satisfies, and thus, we desire that the PBC spectra possess a nonzero spectral area and exhibit spectral reciprocity [62]. Precisely, this indicates that $\sigma^{\mathrm{p}}$ occupies a finite area in the complex frequency plane and $\omega^P(+\boldsymbol{k}) = \omega^P(-\boldsymbol{k})$, where $\boldsymbol{k}$ is the Bloch wavevector. Afterward, the geometric structure should be designed to include macroscopic boundaries that are mismatched with the high-symmetry lines and planes of BZs, resulting in geometry-dependent localization of eigenstates.

We build upon the above understanding by proposing an anisotropic NHM with its permittivity as $\hat{\varepsilon} = \mathrm{diag}(\varepsilon_1, \varepsilon_2)$, where $\varepsilon_1, \varepsilon_2 \in \mathbb{C}$ (Figure 2a). We set the relative magnetic permeability to $1.0$ throughout this work and only consider $H_z$ polarization. Such an NHM respects Lorentz reciprocity ($\hat{\varepsilon}^T = \hat{\varepsilon}$) and thus satisfies spectral reciprocity [74]. Due to the complex nature of $\hat{\varepsilon}$ and the system dimensionality, $\sigma^{\mathrm{p}}$ generally occupies a nonzero spectral area, as shown by the grey regions in Figures 2b and 2c. We utilize the Hamiltonian $H_{\mathrm{CM}}(\boldsymbol{k})$ for the CM, where $\boldsymbol{k} = (k_x, k_y)$, to acquire $\sigma^{\mathrm{p}}$ (see the Methods section for details) [75,76].

We now come to the OBC scenario, which requires the eigenstates to localize according to geometry but not a single $\boldsymbol{\mu}$. We first apply amoeba theory to determine the OBC spectra $\sigma^A$ and the ensuing $\boldsymbol{\mu}$. For our NHMs, we numerically and theoretically verify that $\sigma^A = \sigma^p$ and $\boldsymbol{\mu} = \mathbf{0}$ (see Supporting Information Section 1 for details), indicating that the eigenstates remain extended if following the obtained $\boldsymbol{\mu}$. We next deploy various geometry to showcase the occurrence of eigenstate localizations. To observe such effects, we must require $\varepsilon_1 \neq \varepsilon_2$; otherwise, the NHM becomes isotropic, making all directions equivalent and preventing any macroscopic boundary mismatches. In this context, $\varepsilon_1$ and $\varepsilon_2$ define two optical axes of the NHM, and we then employ the square geometry, one being mismatched with optical axes (red square in Figure 2a) while the other is matched (orange oblique square in Figure 2a). The left panels of Figures 2b and 2c display the OBC spectra of the regular square $\sigma^{\mathrm{mis}}$ (red circles) and oblique square $\sigma^{\mathrm{mat}}$ (orange circles), calculated using the finite element method (FEM) [77]. Clearly, $\sigma^{\mathrm{mat}}$ fills $\sigma^p$, while $\sigma^{\mathrm{mis}}$ shrinks compared with $\sigma^p$. Such spectral stability against geometry implies the possible localization of eigenstates. To illustrate, the top-right panels of Figures 2b and 2c show the eigenstates corresponding to $\omega_3 \in \sigma^{\mathrm{mis}}$ and $\omega_4 \in \sigma^{\mathrm{mat}}$, where $\psi$ is taken as $H_z$ due to polarization. $\psi_4$ fully delocalizes, which obeys the localization predicted by amoeba theory, but $\psi_3$ localizes at the top and bottom boundaries. Apparently, $\psi_3$ is another type of localization state that solely originates from geometry because the PBC spectra for both cases are exactly the same.

To further unveil such localization behavior, we define the layer density of eigenstates (labeled by $v$ with its eigenfrequency $\omega_v$) as follows [66]

$$\Theta_v(r_\perp) = \int \mathrm{d}r_\parallel \left|\psi_v(r_\parallel, r_\perp)\right|^2, \tag{2}$$

where $r_\parallel$ and $r_\perp$ represent the directions parallel and perpendicular to the boundary where the eigenstates localize (Figure 3a). Specifically, for Figure 2b, $r_\parallel = x$ and $r_\perp = y$. Figure 2d displays the corresponding $\Theta_3(y)$ on the semilogarithmic scale, represented by open circles. The dashed black (red) line shows a typical exponential (power-law) decay, which agrees with $\Theta_3(y)$ when near (far from) the boundary at $y = 0$. The inset uses the double logarithmic scale to depict $\Theta_3(y)$ within the dashed box, further confirming the power-law behavior. Such localization is hybrid in nature and clearly distinct from the one described by the conventional GBZ theory. To quantify, we introduce the fractal dimension (FD) for an OBC eigenstate $\psi_v(\boldsymbol{r})$ as [78,79]

$$D[\psi_v] = -\frac{\ln[\sum_r |\psi_v(\boldsymbol{r})|^4]}{\ln \sqrt{N}}, \tag{3}$$

where $\boldsymbol{r} = (r_\parallel, r_\perp)$ are spatial coordinates and $N = L_\parallel \times L_\perp / a_0^2$ is the normalized area of the NHM. If $D[\psi_v]$ equals to 2.0 (1.0), the eigenstate is fully extended (localized at the boundary). The left and right panels of Figure 2e respectively show $D[\psi_v]$ for the OBC eigenstates in Figures 2b and 2c. For the oblique square, $D[\psi_v] = 2.0$, reflecting that all eigenstates are extended. In contrast, for the regular square, $D[\psi_v]$ varies between 1.0 and 2.0, confirming that a significant portion of the eigenstates are hybrid localized. This further validates geometry-dependent localization, thereby realizing ASE in a CM from a phenomenological perspective. Next, we will further demonstrate this by reconstructing the eigenstates in Figures 2b and 2c.

**Solving ASE eigenstates.** Since the CM Hamiltonian $H_{\text{CM}}(\boldsymbol{k})$ is in the reciprocal space even under analytical continuation, while the eigenstates we aim to analyze are in the real space, we shall scrutinize the wave equations governing the NHM for a particular geometry. By considering the Helmholtz equation for the $H_z$ field and discretizing partial differential operators into finite difference forms, we map the NHM to the following LM (see the Methods section for details)

$$H_{\text{OBC}} = \frac{1}{\delta^2} \sum_{n_x,n_y=1}^{N_x,N_y} [(t_x a^\dagger_{n_x+1,n_y} a_{n_x,n_y} + t_y a^\dagger_{n_x,n_y+1} a_{n_x,n_y} + t^+_{xy} a^\dagger_{n_x+1,n_y+1} a_{n_x,n_y}$$
$$+ t^-_{xy} a^\dagger_{n_x+1,n_y-1} a_{n_x,n_y} + h.c.) + u a^\dagger_{n_x,n_y} a_{n_x,n_y}], \tag{4}$$

where $n_x$ and $n_y$ denote the lattice site, $N_x = L_x/\delta$, and $N_y = L_y/\delta$. $\delta$ is the grid size and has been set to $a_0$ in Figure 3. $a_{n_x,n_y}$ ($a^\dagger_{n_x,n_y}$) corresponds to $H_z$ ($H_z^\dagger$) at the $(x = n_x\delta, y = n_y\delta)$ position, and the summation runs over all the spatial coordinates. The hopping parameters and on-site potential in the above Hamiltonian are

$$t_x = -\eta_{yy}, \qquad t_y = -\eta_{xx}, \qquad u = 2(\eta_{xx} + \eta_{yy}), \tag{5}$$

$$t^+_{xy} = \frac{\eta_{xy} + \eta_{yx}}{4}, \qquad t^-_{xy} = -\frac{\eta_{xy} + \eta_{yx}}{4}, \tag{6}$$

$$\begin{pmatrix} \eta_{xx} & \eta_{xy} \\ \eta_{yx} & \eta_{yy} \end{pmatrix} = \frac{1}{\varepsilon_{xx}\varepsilon_{yy} - \varepsilon_{xy}\varepsilon_{yx}} \begin{pmatrix} \varepsilon_{yy} & -\varepsilon_{xy} \\ -\varepsilon_{yx} & \varepsilon_{xx} \end{pmatrix}. \tag{7}$$

Such a discretized Hamiltonian recovers both the spectra and eigenstates calculated using CM and FEM (see Supporting Information Section 2 for details), offering us a tool to investigate the contributions of $\boldsymbol{\mu}$ to the eigenstates.

We now proceed to solve the OBC eigenstates. The ZBCs applied to the above Hamiltonian (eq 4) mean $H_z = 0$ at the boundaries, which are exactly the perfect magnetic conductor (PMC) conditions used throughout this work. To be concrete, we consider a rectangular geometry with $r_\parallel$ and $r_\perp$ being its two orthogonal directions (Figure 3a). If using the spirit of non-Bloch theory, the OBC eigenstates $\psi_\nu(r)$ can be formally expressed as

$$\psi_\nu(r) = \sum_{\beta} A_\beta \beta_\parallel^{r_\parallel/a_0} \beta_\perp^{r_\perp/a_0}, \tag{8}$$

where $\boldsymbol{\beta} = (\beta_\parallel, \beta_\perp)$ are non-Bloch wave vectors and $A_\beta$ are the superposition coefficients. Apparently, the distributions of $A_\beta$ determine the eigenstate behaviors. If $A_\beta$ do not vanish only at a few isolated values of $\boldsymbol{\beta}$, then the eigenstate must be either exponentially localized or fully extended. Therefore, the eigenstates in Figure 2b do not fall into this category. To analyze these eigenstates, we now call the transfer matrix method to solve the eigenstates presented in Figure 2b ($r_\parallel = x$ and $r_\perp = y$) [80]. First, by using the 1D layer for any fixed $y$, we construct the transfer matrix $\mathbb{T}(\omega)$, which connects the $j$-th layer to its adjacent layers (Figure 3a). The eigenvalues of $\mathbb{T}(\omega)$ are the allowed wavevectors $\rho_{y,i}$ ($i = 1, \cdots, 2N_x$) of eigenstates along the $y$-direction (see the Methods section for details). Given $\omega$ and $\rho_{y,i}$, we then enforce the SWC along the $x$-direction (eq 1) and the ChE $\det[H(\beta_x, \beta_y = \rho_{y,i}) - \omega \mathbb{I}] = 0$ to acquire the ensuing $\beta_{x,i}^+$ and $\beta_{x,i}^-$, where $H(\boldsymbol{\beta})$ is the non-Bloch Hamiltonian of eq 4 by applying PBCs and analytical continuation. Once such sets of $\{(\rho_{y,i}, \beta_{x,i}^\pm), i = 1, \cdots, 2N_x\}$ are obtained, the OBC eigenstates defined in eq 8 become

$$\psi_\nu(x, y) = \sum_{i=1}^{2N_x} A_i (\rho_{y,i})^{n_y} [(\beta_{x,i}^+)^{n_x} - (\beta_{x,i}^-)^{n_x}]. \tag{9}$$

The above form automatically satisfies the ZBC along the $x$-direction although there are $2N_x$ standing waves. We finally need to satisfy the ZBC along the $y$-direction. The number of BCs is $2N_x$, which is adequate to determine the $2N_x$ unknown coefficients $A_i$. With details in the Methods section, we arrive at the following matrix equation

$$\mathbb{M}(\omega) \boldsymbol{A} = 0, \tag{10}$$

where $\boldsymbol{A} = (A_1, \cdots, A_{2N_x})^T$ and $\mathbb{M}(\omega)$ is a $2N_x \times 2N_x$ matrix. Solving the null space of $\mathbb{M}(\omega)$ yields $\boldsymbol{A}$, and substituting back to eq 9 will reconstruct the OBC eigenstates.

With the above approach, we calculate the eigenstates corresponding to $\omega_3$ and $\omega_4$, as shown in the bottom-right panels of Figures 2b and 2c, and Figure 2d displays the corresponding $\Theta_3(y)$ by filled circles. Good agreement with the CM results validates the transfer matrix method and offers a tool to digest the distributions of $A_\beta$ and the allowed values of $\boldsymbol{\beta}$. Figures 3b and 3c depict $|A_i|$ in the $(\mu_x, \mu_y)$ and $(k_x, k_y)$ plane, respectively, for the eigenstate in Figure 2b. The circles (lines) are obtained by enforcing SWCs in the $x$- direction ($y$-direction) and matching ZBCs in the $y$-direction ($x$-direction). Clearly, the eigenstate does not originate from a single $\mu$ but rather from a set of $\mu$. The exponential decay near the boundary is dominant by the hot spots of $|A_i|$, as shown in Figures 2d and 3b. The $(k_x, k_y)$ values of such hot spots determine the decay skewness of the eigenstate into the bulk [81]. The power-law behavior arises from the fact that the distribution of $\beta$ in the $\mu$ plane crosses $\mu_x = 0$ or $\mu_y = 0$ (blue dashed lines), leading to the Bloch components contributing to such quasi-long-range behaviors (see Supporting Information Section 3 for details) [66]. For comparison, Figures 3d and 3e present $|A_i|$ for the eigenstate in Figure 2c. Clearly, $|A_i|$ has only one nonzero value at $(\mu_x, \mu_y) = (0,0)$, thus indicating the extended nature of the eigenstate. From the non-Bloch theory viewpoint, such a $\mu$ can be determined by the amoeba theory [72]. However, the set of $\beta$ in Figures 3b and 3c does not align with the predictions of amoeba theory, begging for further investigations.

**ASE mechanisms.** Viewing Figures 3c and 3e, we observe that a single $\mu$ corresponds to isolated values of $k$, while a continuous distribution of $\mu$ also implies the continuous $k$. This distinction reflects the nature of the eigenstates, and thus, the concept of GFS has been introduced [66]. When talking about FS, we mean that an eigenstate with a fixed $\omega \in \sigma^p$ can be expressed by a linear superposition of a specific set of $\psi_k$ within the BZ. Analogously, the GFS implies that an OBC eigenstate with $\omega \in \sigma^A$ can also be expressed as a linear superposition of a specific set of $\psi_\beta$ (eq 8), and Figures 3b and 3c have already unveiled an example. For $d$-dimensional non-Hermitian systems, the dimension of FS is $\dim \text{FS} = d - \dim \sigma^p$, where dim denotes the dimensionality. Concerning our system, $\dim \text{FS} = 0$ implies several isolated Fermi points in BZ (red circles in Figures 3c and 3e), which are rotated with optical axes. If it exists that two Fermi points can project onto the same point in one direction, standing waves can then form in this direction, leading to extended eigenstates. Otherwise, the eigenstates are likely localized, necessitating GFS.

The dimension of GFS is $\dim \text{GFS} = 2d - N_{\text{swc}} - \dim \sigma^A$, where $2d$ is the dimension of $\beta$ and $N_{\text{swc}}$ is the number of SWCs enforced. The key difference is the presence of $N_{\text{swc}}$, which arises from BCs. The contrast between $\dim \text{GFS}$ and $\dim \text{FS}$ reflects the projection of GFS onto the $\mu$ plane. When $N_{\text{swc}} = d$ and $\dim \sigma^p = \dim \sigma^A$, $\dim \text{GFS} = \dim \text{FS}$, indicating either an exponentially decaying or a fully extended eigenstate. Figures 3d and 3e are just the case ($\dim \text{GFS} = \dim \text{FS} = 0$). As sketched in the bottom-left panel of Figure 3a, two SWCs are enforced simultaneously to determine a single $\mu$. In contrast, if either condition fails, $\dim \text{GFS} > \dim \text{FS}$, resulting in coherent interferences of multiple standing waves to satisfy other ZBCs that are not necessarily enforced by SWCs. Figures 3b and 3c belong to this scenario, where $\dim \text{GFS} = 1$, as the GFS forms lines instead of points, as sketched in the bottom-right panel of Figure 3a. For instance, we only enforce the *x*-direction SWC ($N_{\text{swc}} = 1 < d$) at a time, and the coherent superposition of standing waves on the solid blue line then satisfies the *y*-direction ZBC, thus leading to $\dim \text{GFS} = 1$.

Therefore, the criterion for ASE boils down to $\dim \text{GFS} > \dim \text{FS}$ in $d$-dimensional non-Hermitian systems, which has been theoretically and numerically demonstrated in an NHM. We will next present a realistic design by using PhCs to realize this celebrated ASE.

**PhC Implementation.** With the aforementioned understanding and criterion in mind, we employ a PhC to achieve $\varepsilon_1 \neq \varepsilon_2 \in \mathbb{C}$ as the EMT provides a powerful and efficacious recipe by which a CM is adequate to characterize the low-frequency bands. Figure 4a details the designed PhC, with its unit cell being composed of alternating layers of chromium (Cr) and air. The periodic and orthogonal directions are denoted by *1* and *2*, respectively. The utilization of Cr is due to the requirement that imaginary parts of permittivity shall be nonignorable to real parts. We adopt the experimentally determined permittivity for Cr (see Supporting Information Section 4 for details) [82] and then calculate the band structures, $\text{Re}(\omega)$ as a function of $k_1$ and $k_2$, which are shown by filled stars in Figure 4b for several selected $k_2$. Since we focus on the region that both $k_1$ and $k_2$ approach zero, we utilize the EMT [51-54,58], which gives

$$\varepsilon_1 = \frac{\varepsilon_{\text{Cr}}\varepsilon_{\text{Air}}}{(1-f_r)\varepsilon_{\text{Cr}} + f_r\varepsilon_{\text{Air}}}, \qquad \varepsilon_2 = (1-f_r)\varepsilon_{\text{Air}} + f_r\varepsilon_{\text{Cr}}, \qquad (11)$$

where $f_r$ is the filling ratio of Cr. The bands calculated using eq 11 are represented by open circles in Figure 4b. Good agreement validates the EMT, confirming that the condition $\varepsilon_1 \neq \varepsilon_2 \in \mathbb{C}$ has been fulfilled.

We then proceed to demonstrate ASE by deploying the rectangular highlighted by the solid red box in Figure 4a. Its macroscopic boundaries clearly mismatch with $k_1$ and $k_2$, indicating the potential for ASE. For simplicity, if we only aim to realize ASE around 240 THz, we set $\varepsilon_{Cr} = 2.012 + 40.819i$, labeled as the low-frequency approximation (LFA). Based on such approximation, we showcase the calculated eigenstates of PhC and CM around 240 THz, as illustrated in Figure 4c. Figure 4d depicts the layer density $\Theta_\nu(x)$ of PhC and CM, and the dashed grey (purple) lines display the exponential (power-law) decay, confirming the ASE feature. Thus far, we have implemented ASE in a realistic PhC, although we focus on the eigenstates near one frequency. The proposed design methodology and underlying criteria are certainly generalizable, indicating that the ASE realization is not restricted to the parameters presented here.

**Conclusion**

In conclusion, we demonstrate ASE in 2D reciprocal NHMs with anisotropic dielectric tensors, requiring distinct complex permittivity along optical axes and a mismatch between optical axes and geometric structures. Through both numerical and theoretical calculations, we phenomenally demonstrate that ASE exhibits power-law decay instead of exponential decay and physically reveal that the dimensionality of GFS is larger than that of FS when ASE occurs. Finally, we propose a realistic PhC design for achieving ASE, which has been successfully verified. Our work establishes clear criteria for achieving ASE in NHMs and provides a minimal model that is both experimentally feasible and capable of capturing the essential physics. Since ASE enables quasi-long-range interactions, we believe that our recipe is another approach to light manipulation remotely if it can be experimentally demonstrated. This effect, unique to higher-dimensional non-Hermitian systems, demonstrates the potential for broader generalization, including applications in three-dimensional structures [83-86], synthetic dimensions [87-89], or other physical systems [90-96]. Our findings lay the foundations for advancing non-Hermitian photonics and exploring ASE in more complex and practical settings.

**Methods**

**CM Hamiltonian.** To formulate the Hamiltonian for the anisotropic NHM, we express the Helmholtz equation and constitutive relation as

$$\nabla \times \nabla \times \boldsymbol{E}' = \left(\frac{\omega}{c}\right)^2 \hat{\varepsilon} \boldsymbol{E}', \tag{12}$$

$$\boldsymbol{P} = -\frac{\left(\frac{\omega_p}{c}\right)^2}{\left(\frac{\omega}{c}\right)^2 - \left(\frac{\omega_0}{c}\right)^2} \boldsymbol{E}', \tag{13}$$

where $\boldsymbol{E}' = \sqrt{\varepsilon_0}\boldsymbol{E}$. Since the spatial coordinates are rotated by $\theta$ respective to the optical axes, $\hat{\varepsilon}$ formally has four terms $\varepsilon_{xx} = \varepsilon_1 \cos^2\theta + \varepsilon_2 \sin^2\theta$, $\varepsilon_{yy} = \varepsilon_1 \sin^2\theta + \varepsilon_2 \cos^2\theta$, and $\varepsilon_{xy} = \varepsilon_{yx} = (\varepsilon_1 - \varepsilon_2)\cos\theta\sin\theta$. We have only included a single Lorentz response in eq 13 for simplicity, and the generalization to multiple terms is straightforward. For convenience, we define $\omega_p/c = \omega_p'$, $\omega_0/c = \omega_0'$, $\omega/c = \omega'$ and $(\omega_0^2/\omega_p^2)\boldsymbol{P} = \boldsymbol{P}'$, and eqs 12 and 13 can then be written into matrix form [75,76]

$$M_1 \varphi = {\omega'}^2 M_2 \varphi, \qquad \varphi = (E_x' \quad E_y' \quad P_x' \quad P_y')^T, \tag{14}$$

$$M_1 = \begin{pmatrix} k_y^2 + \omega_p'^2 & -k_x k_y & -\omega_p'^2 & 0 \\ -k_x k_y & k_x^2 + \omega_p'^2 & 0 & -\omega_p'^2 \\ -\omega_p'^2 & 0 & \omega_p'^2 & 0 \\ 0 & -\omega_p'^2 & 0 & \omega_p'^2 \end{pmatrix}, \quad (15)$$

$$M_2 = \begin{pmatrix} \varepsilon_1^{(0)} \cos^2\theta + \varepsilon_2^{(0)} \sin^2\theta & (\varepsilon_1^{(0)} - \varepsilon_2^{(0)})\cos\theta\sin\theta & 0 & 0 \\ (\varepsilon_1^{(0)} - \varepsilon_2^{(0)})\cos\theta\sin\theta & \varepsilon_1^{(0)} \sin^2\theta + \varepsilon_2^{(0)} \cos^2\theta & 0 & 0 \\ 0 & 0 & \frac{\omega_p'^2}{\omega_0'^2} & 0 \\ 0 & 0 & 0 & \frac{\omega_p'^2}{\omega_0'^2} \end{pmatrix}. \quad (16)$$

Subsequently, the CM Hamiltonian can be formally represented as

$$H_{\mathrm{CM}} = M_2^{-\frac{1}{2}} M_1 M_2^{-\frac{1}{2}}, \quad (17)$$

where $H_{\mathrm{CM}} \varphi' = \omega'^2 \varphi'$ and $\varphi' = M_2^{\frac{1}{2}} \varphi$.

**LM description to NHMs.** The wave equation of the $H_z$ component in the NHM is

$$(-\eta_{yy}\partial_x^2 - \eta_{xx}\partial_y^2 + \eta_{xy}\partial_x\partial_y + \eta_{yx}\partial_y\partial_x) H_z = \omega'^2 H_z. \quad (18)$$

In order to map to LMs, we express the partial differential operators as the following finite difference forms

$$\partial_x^2 H_z(x,y) = \frac{1}{\delta^2}[H_z(x+\delta,y) + H_z(x-\delta,y) - 2H_z(x,y)], \quad (19)$$

$$\partial_y^2 H_z(x,y) = \frac{1}{\delta^2}[H_z(x,y+\delta) + H_z(x,y-\delta) - 2H_z(x,y)], \quad (20)$$

$$\partial_x \partial_y H_z(x,y) = \frac{1}{4\delta^2}[H_z(x+\delta,y+\delta) + H_z(x-\delta,y-\delta) - H_z(x+\delta,y-\delta) \\ - H_z(x-\delta,y+\delta)]. \quad (21)$$

Substituting eqs 19–21 into eq 18 will obtain $H_{\mathrm{OBC}}$ (eq 4).

**Transfer matrix method.** We consider the case shown in Figure 3a and rewrite $H_{\mathrm{OBC}}$ in eq 4 as the interaction between adjacent layers in the $y$ direction

$$H_{\mathrm{OBC}} = \frac{1}{\delta^2} \sum_{l=0,\pm1} \sum_{n_y=1}^{N_y} \left( \boldsymbol{a}_{n_y}^\dagger \hat{h}_l \boldsymbol{a}_{n_y+l} \right), \quad (22)$$

where $\boldsymbol{a}_{n_y}^\dagger = (a_{n_x=1,n_y}^\dagger, \cdots, a_{n_x=N_x,n_y}^\dagger)$ represents the generation operator of the $j$-th layer

$$\hat{h}_{-1} \phi_{n_y-1} + (\hat{h}_0 - \omega'^2 \delta^2 \mathbb{I}_{N_x}) \phi_{n_y} + \hat{h}_1 \phi_{n_y+1} = 0, \quad (23)$$

where $\hat{h}_{0,\pm 1}$ are the hopping strengths between different layers and $\phi_{n_y}$ is expressed as $(\psi_{x=\delta,n_y\delta},\cdots,\psi_{x=L_x,n_y\delta})^T$. The ZBCs in the $y$ direction are $\phi_{n_y=0}=0$ and $\phi_{n_y=N_y+1}=0$, and thus, when $n_y=1$ and $n_y=N_y$, eq 23 becomes BCs required to satisfy. By writing eq 23 in the matrix form, we can obtain the layer transfer matrix $\mathbb{T}(\omega)$ in the real space

$$\mathbb{T}(\omega) = \begin{pmatrix} \hat{h}_1^{-1}(\omega'^2\delta^2 \mathbb{I}_{N_x} - \hat{h}_0) & -\hat{h}_1^{-1}\hat{h}_{-1} \\ \mathbb{I}_{N_x} & 0 \end{pmatrix}, \quad (24)$$

which is a $2N_x \times 2N_x$ square matrix. Its corresponding non-Bloch layer transfer matrix $\mathbb{T}(\beta_x,\omega)$ is

$$\mathbb{T}(\beta_x,\omega) = \begin{pmatrix} \dfrac{(\omega'^2\delta^2 - h_0(\beta_x))}{h_1(\beta_x)} & \dfrac{-h_{-1}(\beta_x)}{h_1(\beta_x)} \\ 1 & 0 \end{pmatrix}, \quad (25)$$

where $h_0(\beta_x) = t_x(\beta_x + \beta_x^{-1}) + u$ and $h_{\pm 1}(\beta_x) = t_{xy\pm}\beta_x + t_{xy\mp}\beta_x^{-1} + t_y$. Thus, the solutions of $\beta_x$ can be obtained by the following equation

$$\det[\mathbb{T}(\beta_x,\omega) - \rho_y \mathbb{I}] = 0. \quad (26)$$

Only solutions of $\boldsymbol{\beta}$ that satisfy both eq 26 and the GBZ condition (eq 1) constitute the non-Bloch basis vectors of the eigenstate.

**Acknowledgments**

We thank Dr. Kai Zhang for the helpful discussions. This work is supported by the National Key R&D Program of China (No. 2022YFA1404701, No. 2022YFA1404500), the National Natural Science Foundation of China (No. 12174072, No. 2021hwyq05, No.12347144).

**References**

[1] J. B. Pendry, Negative refraction makes a perfect lens, Phys. Rev. Lett. **85**, 3966 (2000).

[2] N. Yu and F. Capasso, Flat optics with designer metasurfaces, Nat. Mater. **13**, 139 (2014).

[3] J. B. Pendry, A. J. Holden, D. J. Robbins, and W. J. Stewart, Low frequency plasmons in tine-wire structures, J. Phys.: Condens. Matter **10**, 4785 (1998).

[4] N. Fang, H. Lee, C. Sun, and X. Zhang, Sub-diffraction-limited optical imaging with a silver superlens, Science **308**, 534 (2005).

[5] X. Zhang and Z. Liu, Superlenses to overcome the diffraction limit, Nat. Mater. **7**, 435 (2008).

[6] S. B. Glybovski, S. A. Tretyakov, P. A. Belov, Y. S. Kivshar, and C. R. Simovski, Metasurfaces: from microwaves to visible, Physics Reports **634**, 1 (2016).

[7] I. Liberal and N. Engheta, Near-zero refractive index photonics, Nat. Photonics **11**, 149 (2017).

[8] C. M. Watts, X. Liu, and W. J. Padilla, Metamaterial electromagnetic wave absorbers, Adv. Mater. **24**, OP98 (2012).

[9] R. A. Shelby, D. R. Smith, and S. Schultz, Experimental verification of a negative index of refraction, Science **292**, 77 (2001).

[10] E. Arbabi, A. Arbabi, S. M. Kamali, Y. Horie, M. Faraji-Dana, and A. Faraon, MEMS-tunable dielectric metasurface lens, Nat. Commun. **9**, 812 (2018).

[11] S. Ma, B. Yang, and S. Zhang, Topological photonics in metamaterials, Photonics insights **1**, R02 (2022).

[12] N. Yu, P. Genevet, M. A. Kats, F. Aieta, J.-P. Tetienne, F. Cappsso, and Z. Gaburro, Light propagation with phase discontinuities: Generalized laws of reflection and refraction, Science **334**, 333 (2011).

[13] S. Sun, Q. He, S. Xiao, Q. Xu, X. Li, and L. Zhou, Gradient-index meta-surfaces as a bridge linking propagating waves and surface waves, Nat. Mater. **11**, 426 (2012).

[14] Z.-K. Shao, H.-Z. Chen, S. Wang, X.-R. Mao, Z.-Q. Yang, S.-L. Wang, X.-X. Wang, X. Hu, and R.-M. Ma, A high-performance topological bulk laser based on band-inversion-induced reflection, Nat. Nanotechnol. **15**, 67 (2020).

[15] W. Zhang, X. Xie, H. Hao, J. Dang, S. Xiao, *et al.*, Low-threshold topological nanolasers based on the second-order corner, Light Sci. Appl. **9**, 109 (2020).

[16] Q. Chen, L. Zhang, F. Chen, Q. Yan, R. Xi, H. Chen, and Y. Yang, Photonic topological valley-locked waveguides, ACS Photonics **8**, 1400 (2021).

[17] L. Yu, H. Xue, R. Guo, E. Chan, Y. Tern, C. Soci, B. Zhang, and Y. D. Chong, Dirac mass induced by optical gain and loss, Nature **632**, 63 (2024).

[18] T. Dai, A. Ma, J. Mao, Y. Ao, X. Jia, *et al.*, A programmable topological photonic chip, Nat. Mater. **23**, 928 (2024).

[19] S. Mandal, G.-G. Liu, and B. Zhang, Topology with memory in nonlinear driven-dissipative photonic lattices, ACS Photonics **10**, 147 (2023).

[20] Z.N. Liu, X. Q. Zhao, J. Yao, C. Zhang, J. L. Xu, S. Zhu, and H. Liu, Designing a transition photonic band with a Moire synthetic sphere, Phys. Rev. Appl. **19**, 044054 (2023).

[21] L. Huang, W. Zhang, and X. Zhang, Moire quasibound states in the continuum, Phys. Rev. Lett. **128**, 253901 (2022).

[22] X. Huang, Y. Lai, Z. Hang, H. Zheng, and C. T. Chan, Dirac cones induced by accidental degeneracy in photonic crystals and zero-refractive-index materials, Nat. Mater. **10**, 582 (2011).

[23] C. Liu, Q. Ma, Z. J. Luo, Q. R. Hong, Q. Xiao, *et al.*, A programmable diffractive deep neural network based on a digital-coding metasurface array, Nat. Electron. **5**, 113 (2022).

[24] W. Song, T. Li, S. Wu, Z. Wang, C. Chen, *et al.*, Dispersionless coupling among optical waveguides by artificial gauge field, Phys. Rev. Lett. **129**, 053901 (2022).

[25] S. Wang, P. C. Wu, W.-C. Su, Y.-C. Lai, H. C. Cheng, *et al.*, Broadband achromatic optical metasurface devices, Nat. Commun. **8**, 187 (2017).

[26] J. Y. Dai, L. X. Yang, J. C. Ke, M. Z. Chen, W. Tang, *et al.*, High-efficiency synthesizer for spatial waves based on space-time-coding digital metasurface, Laser & Photonics Reviews **14**, 1900133 (2020).

[27] Y. Pan, C. Cui, Q. Chen, F. Chen, L. Zhang, *et al.*, Real higher-order Weyl photonic crystal, Nat. Commun. **14**, 6636 (2023).

[28] D. Li, B. Gao, H. Ma, W.-Y. Yin, H. Chen, and H. Qian, Ultrafast tunable scattering of optical antennas driven by metallic quantum wells, ACS Photonics **9**, 2346 (2022).

[29] X. Zhang, J. Chen, R. Chen, C. Wang, T. Cai, R. Abdi-Ghaleh, H. Chen, and X. Lin, Perspective on meta-boundaries, ACS Photonics **10**, 2102 (2023).

[30] C. Qian, X. Lin, Y. Yang, F. Gao, Y. Shen, *et al.*, Multifrequency superscattering from subwavelength hyperbolic structures, ACS Photonics **5**, 1506 (2018).

[31] K. Ding, C. Fang, and G. Ma, Non-Hermitian topology and exceptional-point geometries, Nat. Rev. Phys. **4**, 745 (2022).

[32] Y. Ashida, Z. Gong, and M. Ueda, Non-Hermitian physics, Advances in Physics **69**, 249 (2020).

[33] E. Bergholtz, J. Budich, and F. Kunst, Exceptional topology of non-Hermitian systems, Rev. Mod. Phys. **93**, 015005 (2021).

[34] R. Lin, T. Tai, L. Li, and C. H. Lee, Topological non-Hermitian skin effect, Frontiers of Physics **18**, 53605 (2023).

[35] N. Okuma and M. Sato, Non-Hermitian topological phenomena: A review, Annual Review **14**, 83 (2023).

[36] M.-A. Miri and A. Alu, Exceptional points in optics and photonics, Science **363**, eaar7709 (2019).

[37] K. Ding, G. Ma, M. Xiao, Z. Q. Zhang, and C. T. Chan, Emergence, coalescence, and topological properties of multiple exceptional points and their experimental realization, Phys. Rev. X **6**, 021007 (2016).

[38] H. Zhou, C. Peng, Y. Yoon, C. W. Hsu, K. A. Nelson, L. Fu, J. D. Joannopoulos, M. Soljacic, and B. Zhen, Observation of bulk Fermi arc and polarization half charge from paired exceptional points, Science **359**, 1009 (2018).

[39] T. Liu, S. An, Z. Gu, S. Liang, H. Gao, G. Ma, and J. Zhe, Chirality-switchable acoustic vortex emission via non-Hermitian selective excitation at an exceptional point, Science Bulletin **67**, 1131 (2012).

[40] H. Z. Chen, T. Liu, H.-Y. Luan, R.-J. Liu, X.-Y. Wang, *et al.*, Revealing the missing dimension at an exceptional point, Nature Physics **16**, 571 (2020).

[41] S. Weidemann, M. Kremer, T. Helbig, T. Hofmann, A. Stegmaier, M. Greiter, R. Tonmale, and A. Szameit, Topological funneling of light, Science **368**, 311 (2020).

[42] A. Ghatak, M. Brandenbourger, J. van Wezel, and C. Coulais, Observation of non-Hermitian topology and its bulk-edge correspondence in an active mechanical metamaterial, Proceedings of the National Academy of Sciences **117**, 29561 (2020).

[43] L. Xiao, T. Deng, K. Wang, G. Zhu, Z. Wang, W. Yi, and P. Xue, Non-Hermitian bulk-boundary correspondence in quantum dynamics, Nature Physics **16**, 761 (2020).

[44] S. Yao and Z. Wang, Edge states and topological invariants of non-Hermitian systems, Phys. Rev. Lett. **121**, 086803 (2018).

[45] K. Yokomizo and S. Murakami, Non-Bloch band theory of non-Hermitian systems, Phys. Rev. Lett. **123**, 066404 (2019).

[46] K. Zhang, Z. Yang, and C. Fang, Correspondence between winding numbers and skin modes in non-Hermitian Systems, Phys. Rev. Lett. **125**, 126402 (2020).

[47] Z. Yang, K. Zhang, C. Fang, and J. Hu, Non-Hermitian bulk-boundary correspondence and auxiliary generalized Brillouin zone theory, Phys. Rev. Lett. **125**, 226402 (2020).

[48] W.-T. Xue, M.-R. Li, Y.-M. Hu, F. Song, and Z. Wang, Simple formulas of directional amplification from non-Bloch band theory, Phys. Rev. B **103**, L241408 (2021).

[49] L. Li, S. Mu, C. Lee, and J. Gong, Quantized classical response from spectral winding topology, Nat. Commun. **12**, 5294 (2021).

[50] C. C. Wanjura, M. Brunelli, and A. Nunnenkamp, Topological framework for directional amplification in driven-dissipative cavity arrays, Nat. Commun. **11**, 3149 (2020).

[51] P. Sheng, Introduction to Wave Scattering, Localization and Mesoscopic Phenomena, Springer: Berlin, Heidelberg, 2006.

[52] X. Cui, K. Ding, J.-W. Dong, and C. T. Chan, Realization of complex conjugate media using non-PT-symmetric photonic crystals, Nanophotonics **9**, 195 (2020).

[53] Y. Wu, J. Li, Z.-Q. Zhang, and C. T. Chan, Effective medium theory for magnetodielectric composites: Beyond the long-wavelength limit, Phys. Rev. B **74**, 085111 (2006).

[54] X. Huang, Y. Lai, Z. Hang, H. Zheng, and C. T. Chan, Dirac cones induced by accidental degeneracy in photonic crystals and zero-refractive-index materials, Nat. Mater. **10**, 582 (2011).

[55] Y.-M. Hu, Y.-Q. Huang, W.-T. Xue, and Z. Wang, Non-Bloch band theory for non-Hermitian continuum systems, Phys. Rev. B **110**, 205429 (2024).

[56] H. Ding and K. Ding, Non-Bloch theory for spatiotemporal photonic crystals assisted by continuum effective medium, Phys. Rev. Res. **6**, 033167 (2024).

[57] K. Yokomizo, T. Yoda, and S. Murakami, Non-Hermitian waves in a continuous periodic model and application to photonic crystals, Phys. Rev. Res. **4**, 023089 (2022).

[58] T. Yoda, Y. Moritake, K. Takata, K. Yokomizo, S. Murakami, and M. Notomi, Optical non-Hermitian skin effect in two-dimensional uniform media, arXiv:2303.05185.

[59] K. Yokomizo, T. Yoda, and Y. Ashida, Non-Bloch band theory of generalized eigenvalue problems, Phys. Rev. B **109**, 115115 (2024).

[60] A. Poddubny, J. Zhong, and S. Fan, Mesoscopic non-Hermitian skin effect, Phys. Rev. A **109**, L061504 (2024).

[61] K. Kawabata, K. Shiozaki, M. Ueda, and M. Sato, Symmetry and topology in non-Hermitian physics, Phys. Rev. X **9**, 041015 (2019).

[62] K. Zhang, Z. Yang, and C. Fang, Universal non-Hermitian skin effect in two and higher dimensions, Nat. Commun. **13**, 2496 (2022).

[63] W. Wang, M. Hu, X. Wang, G. Ma, and K. Ding, Experimental realization of geometry-dependent skin effect in a reciprocal two-dimensional lattice, Phys. Rev. Lett. **131**, 207201 (2023).

[64] Q. Zhou, J. Wu, Z. Pu, J. Lu, X. Huang, W. Deng, M. Ke, and Z. Liu, Observation of geometry-dependent skin effect in non-Hermitian phononic crystals with exceptional points, Nat. Commun. **14**, 4569 (2023).

[65] T. Wan, K. Zhang, J. Li, Z. Yang, and Z. Yang, Observation of the geometry-dependent skin effect and dynamical degeneracy splitting, Science Bulletin **68**, 2330 (2023).

[66] K. Zhang, C. Shu, and K. Sun, Algebraic non-Hermitian skin effect and unified non-Bloch band theory in arbitrary dimensions, arXiv:2406.06682.

[67] C. Shu, K. Zhang, and K. Sun, Ultra spectral sensitivity and non-local bi-impurity bound states from quasi-long-range non-Hermitian skin modes, arXiv:2409.13623.

[68] Y. Xiong, Z.-Y. Xing, and H. Hu, Non-Hermitian skin effect in arbitrary dimensions: non-Bloch band theory and classification, arXiv:2407.01296.

[69] F. Song, H.-Y. Wang, and Z. Wang, Fragile non-Bloch spectrum and unconventional Green's function, arXiv:2410.23175.

[70] D. Nakamura, T. Bessho, and M. Sato, Bulk-boundary correspondence in point-gap topological phases, Phys. Rev. Lett.**132**, 136401 (2024).

[71] K. Kawabata, T. Bessho, and M. Sato, Classification of exceptional points and non-Hermitian topological semimetals, Phys. Rev. Lett. **123**, 066405 (2019).

[72] H.-Y. Wang, F. Song, and Z. Wang, Amoeba formulation of non-Bloch band theory in arbitrary dimensions, Phys. Rev. X **14**, 021011 (2024).

[73] R. Zheng, J. Lin, J. Liang, K. Ding, J. Lu, W. Deng, M. Ke, X. Huang, and Z. Liu, Experimental probe of point gap topology from non-Hermitian Fermi-arcs, Communications Physics **7**, 298 (2024).

[74] X. Cui, K. Ding, J.-W. Dong, and C. T. Chan, Exceptional points and their coalescence of -symmetric interface states in photonic crystals, Phys. Rev. B **100**, 115412 (2019).

[75] A. Raman and S. Fan, Photonic band structure of dispersive metamaterials formulated as a Hermitian eigenvalue problem, Phys. Rev. Lett. **104**, 087401 (2010).

[76] D. Wang, Y. Wu, Z. Q. Zhang, and C. T. Chan, Non-Abelian frame charge flow in photonic media, Phys. Rev. X **13**, 021024 (2023).

[77] COMSOL Multiphysics 3.5, developed by COMSOL ©, network license (2008).

[78] Y. Qin, K. Zhang, and L. Li, Geometry-dependent skin effect and anisotropic Bloch oscillations in a non-Hermitian optical lattice, Phys. Rev. A **109**, 023317 (2024).

[79] N. Mace, F. Alet, and N. Laflorencie, Multifractal scalings across the many-body localization transition, Phys. Rev. Lett. **123**, 180601 (2019).

[80] F. K. Kunst and V. Dwivedi, Non-Hermitian systems and topology: A transfer-matrix perspective, Phys. Rev. B **99**, 245116 (2019).

[81] K. Zhang, Z. Yang, and K. Sun, Edge theory of non-Hermitian skin modes in higher dimensions, Phys. Rev. B **109**, 165127 (2024).

[82] A. D. Rakic, A. B. Djurisic, J. M. Elazar, and M. L. Majewski, Optical properties of metallic films for vertical-cavity optoelectronic devices, Applied Optics **37**, 5271 (1998).

[83] E. Armstrong and C. O'Dwyer, Artificial opal photonic crystals and inverse opal structures -fundamentals and applications from optics to energy storage, Journal of Materials Chemistry C **3**, 6109 (2015).

[84] D. Wei, C. Wang, H. Wang, X. Hu, D. Wei, X. Fang, Y. Zhang, D. Wu, Y. Hu, J. Li, S. Zhu, and M. Xiao, Experimental demonstration of a three-dimensional lithium niobate nonlinear photonic crystal, Nat. Photonics **12**, 596 (2018).

[85] L. Lu, Z. Wang, D. Ye, L. Ran, L. Fu, J. D. Joannopoulos, and M. Soljacic, Experimental observation of Weyl points, Science **349**, 622 (2015).

[86] B. Yang, Q. Guo, B. Tremain, R. Liu, L. E. Barr, *et al.*, Ideal Weyl points and helicoid surface states in artificial photonic crystal structures, Science **359**, 1013 (2018).

[87] Q. Lin, M. Xiao, L. Yuan, and S. Fan, Photonic Weyl point in a two-dimensional resonator lattice with a synthetic frequency dimension, Nat. Commun. **7**, 13731 (2016).

[88] A. Dutt, Q. Lin, L. Yuan, M. Minkov, M. Xiao, and S. Fan, A single photonic cavity with two independent physical synthetic dimensions, Science **367**, 59 (2020).

[89] L. Yuan, A. Dutt, and S. Fan, Synthetic frequency dimensions in dynamically modulated ring resonators, APL Photonics **6**, 071102 (2021).

[90] J. Lu, W. Deng, X. Huang, M. Ke, and Z. Liu, Non-Hermitian topological phononic metamaterials, Advanced Materials, 2307998 (2023).

[91] X. Wang, X. Fang, D. Mao, Y. Jing, and Y. Li, Extremely asymmetrical acoustic metasurface mirror at the exceptional point, Phys. Rev. Lett. **123**, 214302 (2019).

[92] B. Hu, Z. Zhang, Y. Liu, D. Liao, Y. Zhu, H. Zhang, Y. Cheng, X. Liu, and J. Christensen, Engineering higher-order topological confinement via acoustic non-Hermitian textures, Advanced Materials **36**, 2406567 (2024).

[93] C. Scheibner, William T. M. Irvine, and V. Vitelli, Non-Hermitian band topology and skin modes in active elastic media, Phys. Rev. Lett. **125**, 118001 (2020).

[94] Y. Chen, X. Li, C. Scheibner, V. Vitelli, and G. Huang, Realization of active metamaterials with odd micropolar elasticity, Nat. Commun. **12**, 5935 (2021).

[95] E. Zhao, Z. Wang, C. He, Ting Fung Jeffrey Poon, K. K. Pak, Y.- J. Liu, P. Ren, X.-J. Liu, and G.-B. Jo, Two-dimensional non-Hermitian skin effect in an ultracold Fermi gas, Nature **637**, 565 (2025).

[96] W. Wang, X. Wang, and G. Ma, Non-Hermitian morphing of topological modes, Nature **608**, 50 (2022).

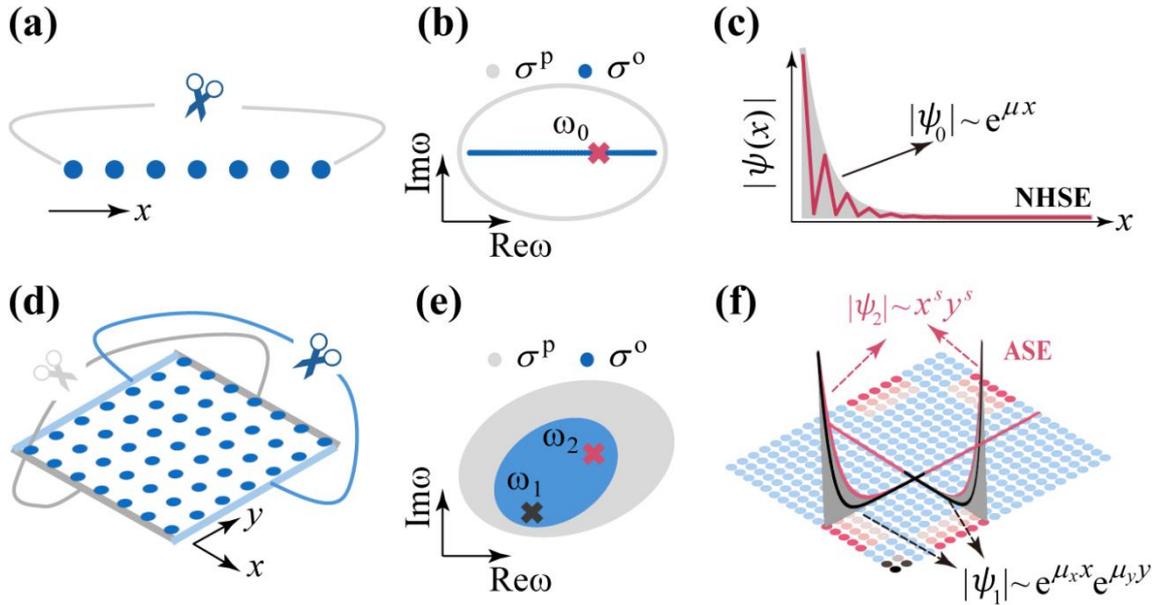

**Figure 1.** Schematic representation of exponentially localized and power-law localized skin effects. (a-c) Illustration of a typical (a) 1D non-Hermitian LM, (b) its spectrum, and (c) eigenstate. The PBC (OBC) configuration and corresponding spectra are sketched in grey (blue). A generic eigenstate corresponding to one selected frequency $\omega_0$ on the OBC spectrum [marked by a red cross in (b)] is shown in (c), with its envelope highlighted by the grey shadow. (d-f) One representative geometry configuration of a 2D non-Hermitian system. A 2D non-Hermitian LM is placed (d) in the rectangular geometry with (e) its typical spectra and (f) eigenstates. The grey (blue) area in (e) indicates the PBC (OBC) spectrum. For $\omega_1$ (black cross) and $\omega_2$ (red cross) in the OBC spectrum, the eigenstate envelopes depicted in (f) exhibit exponentially localized (black) and power-law localized (red) characteristics in the $x$ and $y$ directions.

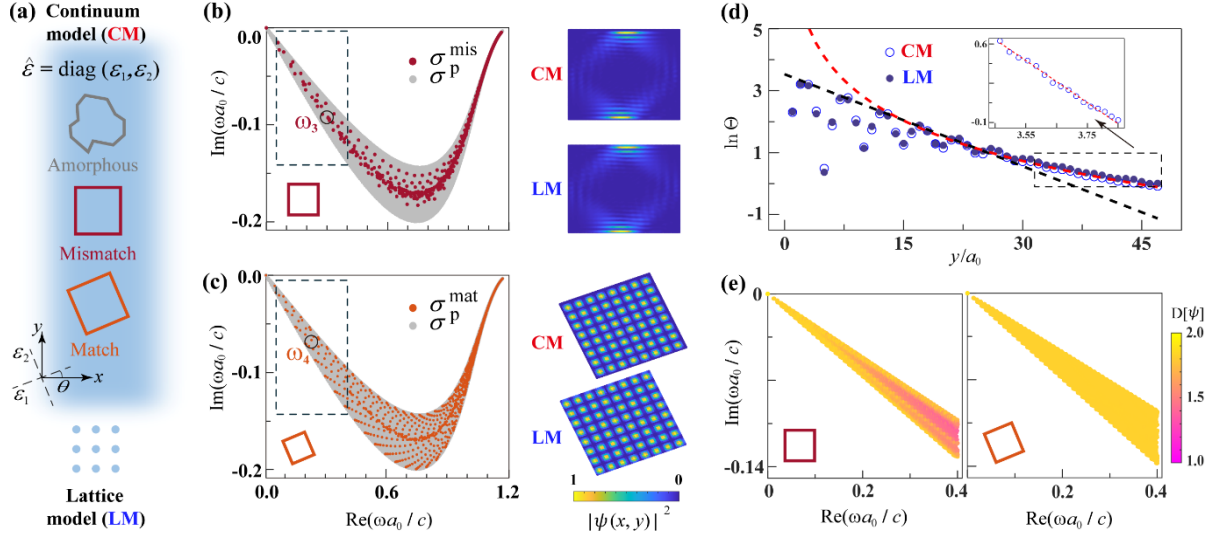

**Figure 2.** Spectra and eigenstates of NHMs with different geometric structures. (a) Schematic of a non-Hermitian anisotropic metamaterial with optical axes labeled as *1* and *2* (dashed lines). Three different geometric structures—amorphous, square, and oblique square—are highlighted in grey, red, and orange, respectively. The bottom panel illustrates the discretization process for mapping the NHM to an LM. (b,c) OBC spectra (left) and typical eigenstates (right) for the regular (b) and oblique square (c) geometries. The grey regions in (b,c) represent PBC spectra $\sigma^p$ of the NHM. The top and bottom panels on the right, respectively, show the eigenstates marked on the left calculated from CM and LM. (d) Layered density $\Theta_\nu(r_\perp)$ for the eigenstate at $\omega_3$ along the $y$-direction. Open and filled circles represent the results from CM and LM, respectively. The black (red) dashed line depicts an exponential (power-law) decay function $\Theta(y)\sim e^{\mu y}$ ($\sim y^s$) with parameters $\mu a_0 = -0.099$ ($s = -1.883$). The inset shows $\Theta_\nu(r_\perp)$ in the double logarithmic scales. (e) FD $D[\psi_\nu]$ of OBC eigenstates for the NHM in the regular (left) and oblique (right) square geometry. Only low-frequency portions of the OBC spectra are shown, highlighted by the dashed boxes in (b) and (c). The permittivities in use are $\varepsilon_1(\omega) = \varepsilon_1^{(0)} - \frac{\omega_p^2}{\omega^2-\omega_0^2}$ and $\varepsilon_2(\omega) = \varepsilon_2^{(0)} - \frac{\omega_p^2}{\omega^2-\omega_0^2}$, where $\varepsilon_1^{(0)} = 8.222 + 6i$, $\varepsilon_2^{(0)} = 4.222 + 6i$, $\omega_p a_0/c = 2$, $\omega_0 a_0/c = 1.2$, and $a_0 = 1$ μm. $c$ is the speed of light in the vacuum. The geometric parameters used are $L_\parallel = L_\perp = 100 a_0$ and $\theta = \pi/8$. The selected frequencies in (b,c) are $\omega_3 a_0/c = 0.310 - 0.093i$ and $\omega_4 a_0/c = 0.220 - 0.061i$.

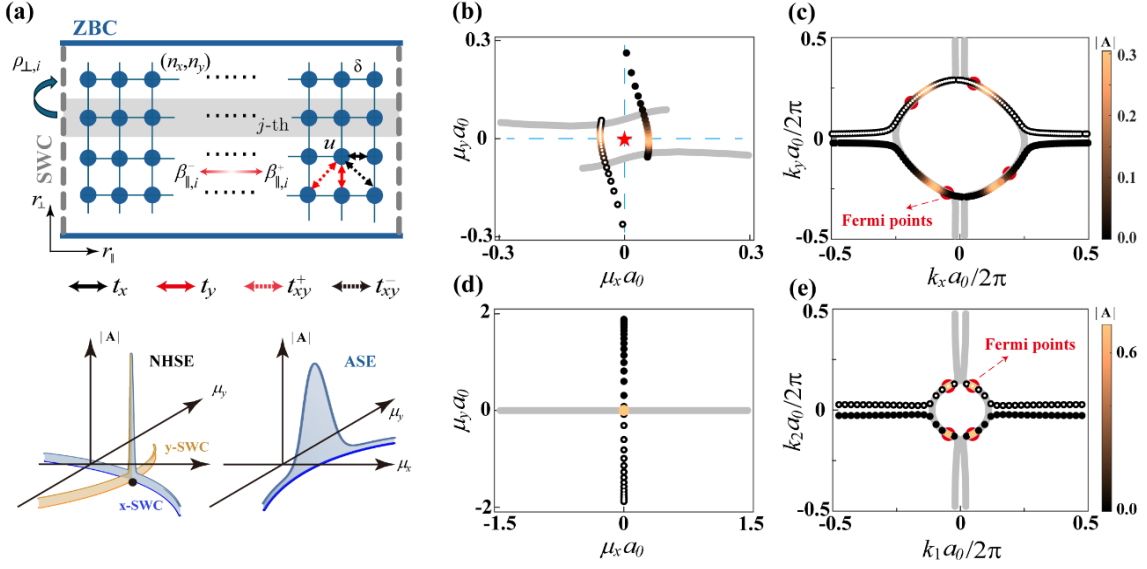

**Figure 3.** Mechanism of ASE and GFS. (a) Upper: The schematic diagram of solving the eigenstates by the transfer matrix method. The NHM is discretized into an LM with blue dots denoting lattice sites $(n_x, n_y)$. $\beta_{\parallel,i}^+$ and $\beta_{\parallel,i}^-$ represent a pair of wave vectors that satisfy SWC in the $r_\parallel$ direction and couple with the propagation factor $\rho_{\perp,i}$ in the $r_\perp$ direction. Lower: Diagram illustrating the GFS by enforcing different SWCs (blue and orange lines) and the corresponding weighting coefficients (solid lines) for exponentially localized (left panel) and power-law localized (right panel). (b-e) GFS in the $(\mu_x, \mu_y)$ and $(k_x, k_y)$ planes. The color scales indicate the weighting coefficients on GFS used to reconstruct the corresponding eigenstates. The results in (b,c) and (d,e) correspond to the eigenstate at $\omega_3$ and $\omega_4$. The circles and lines are the GFS by enforcing the SWC in the $x$ and $y$ direction, and the pentagram represents the position of $\boldsymbol{\mu}$ calculated by amoeba theory. The red circles in (c,e) represent Fermi points.
22

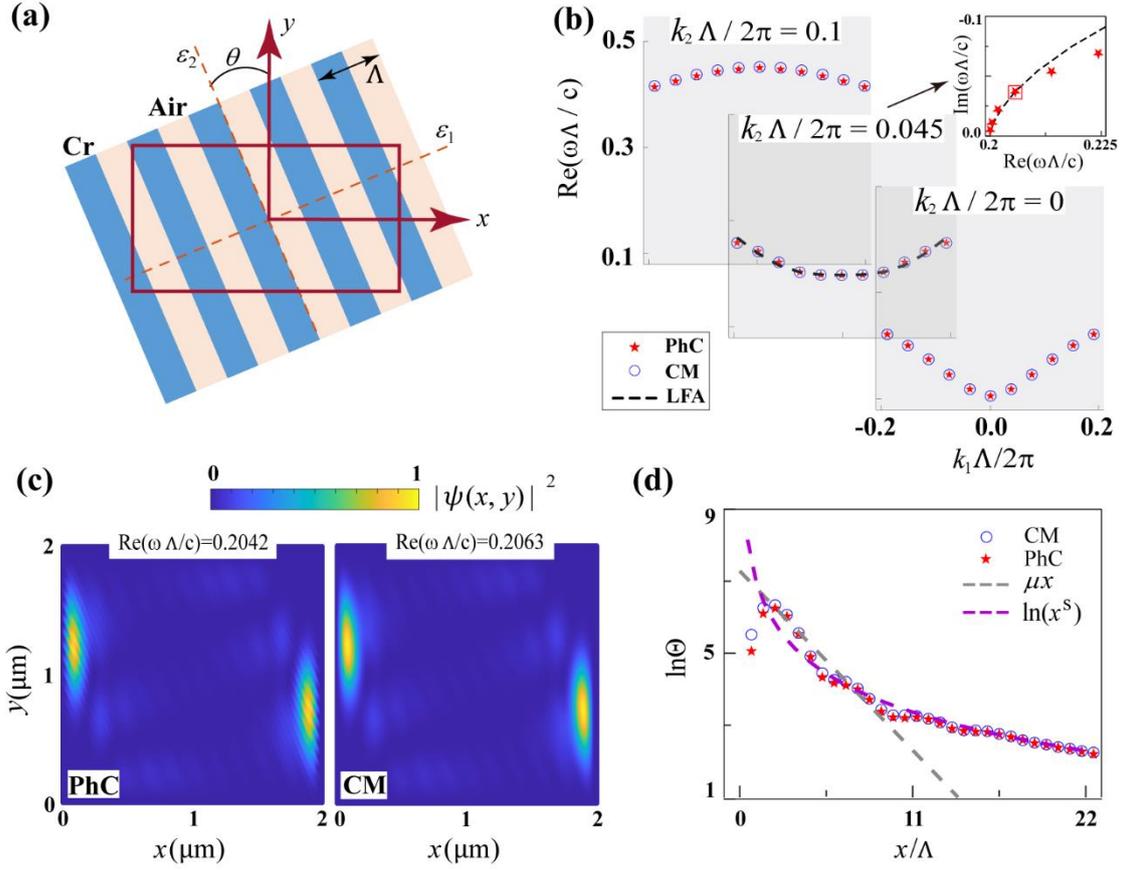

**Figure 4.** Realization of ASE by PhCs. (a) A metamaterial composed of multilayers of Cr and air. The period of metamaterials is $\Lambda$. The optical axes (coordinates) are labeled as $\varepsilon_1$ and $\varepsilon_2$ ($x$ and $y$), and the angle between the axis $2$ and the $y$-axis is $\theta$. (b) PBC bands $\text{Re}[\omega\Lambda/c]$ as a function of $k_1\Lambda/2\pi$ for $k_2\Lambda/2\pi = 0$, $0.045$, and $0.1$ of the PhCs (filled stars) and CMs (open circles). The black dashed line is calculated under LFA, and the inset compares the PBC spectra of LFA and PhC when $k_2\Lambda/2\pi = 0.045$. (c) $|\psi(x,y)|^2$ of a chosen OBC eigenstate for the PhC (left panel, $243.78 - 96.459i$ THz) and the CM (right panel, $246.30 - 97.994i$ THz) under PMCs. The side length of the square is 2000 nm. (d) Layer density $\Theta_\nu(x)$ for the PhC (filled stars) and the CM (open circles) in (c). The grey (purple) dashed line represents an exponential (power-law) decay function $\Theta(x) \sim e^{\mu x}$ ($\sim x^s$) with parameters $\mu\Lambda = -0.453$ ($s = -1.549$). The used geometry parameters are $\Lambda = 40$ nm, $f_r = 0.5$, and $\theta = \pi/8$. The permittivity of Cr is taken from Ref. [82].